# A Protocol-Agnostic Backscatter-Based Security Layer for Ultra-Low-Power SWIPT IoT Networks

Taki E. Djidjekh, Member, IEEE, Alexandru Takacs, Member, IEEE, Gaël Loubet, Member, IEEE, Lamoussa Sanogo, and Daniela Dragomirescu, Senior Member, IEEE

*Abstract*—This paper presents a lightweight, protocol-agnostic security enhancement for Simultaneous Wireless Information and Power Transfer (SWIPT) in Internet of Things (IoT) applications. Building on a backscatter-based identification mechanism, the proposed approach introduces a secure, energy-efficient layer that operates independently of communication protocols and with minimal hardware modification. A rectifier-driven backscattering scheme embedded in battery-free sensing nodes enables authentication without activating conventional RF transceivers, thereby reducing power consumption while ensuring secure device identification. To assess robustness, replay attacks are emulated on standard LoRaWAN Activation By Personalization (ABP) encryption, highlighting vulnerabilities and demonstrating the relevance of the proposed solution. The approach is experimentally validated in a real Wireless Sensor Network (WSN) using LoRaWAN-compatible, battery-free sensing nodes equipped with compact, low-profile antennas, confirming both practicality and scalability for space-constrained IoT deployments. Results show that the method achieves secure identification, reliable energy harvesting, and data transmission with negligible impact on node autonomy. The proposed approach offers a practical, energy-efficient, and scalable security framework for SWIPT-enabled IoT systems, strengthening device authentication without altering existing communication protocols or compromising power autonomy.

## I. INTRODUCTION

THE Internet of Things (IoT) is witnessing an exponential increase in the use of wireless nodes operating in Cyber-Physical Systems (CPS) for diverse applications such as Structural Health Monitoring (SHM), bio-medical and healthcare, autonomous driving systems, and beyond. The deployment of Wireless Sensor Networks (WSNs) at such a large scale introduces two major antagonistic challenges: security and energy autonomy.

Security challenges in IoT systems are widely addressed through protocol-level solutions, primarily based on software cryptographic techniques. While these methods provide robust protection against threats such as non-authorized intrusion, replay and relay attacks, and eavesdropping, they impose significant computational overhead [1], [2]. On the other hand, energy autonomy and efficiency are achieved through the minimization of energy consumption in wireless sensing nodes and the use of energy harvesting and Wireless Power Transfer (WPT) techniques [3]. Recently, the design of battery-free and energy-autonomous wireless sensing nodes has become a promising area of research, with devices powered by far-field WPT and communicating using lightweight wireless technologies such as LoRaWAN and Bluetooth Low Energy (BLE) [4].

Despite advances in both security and energy autonomy, IoT systems remain vulnerable to attacks such as replay and man-in-the-middle due to the constraints of traditional protocols [5], [6]. Addressing these security issues often demands additional computational resources, leading to increased energy usage. As a result, lightweight cryptographic techniques [7], [8] have been proposed to address security in resource-constrained IoT devices. While these methods reduce computational and energy demands, they face scalability issues and trade-offs between security and energy efficiency. Other alternatives, such as Secure Elements (SE) [9], Intrusion Detection Systems (IDS) [10], [11], and multilayer signatures [12], also come with additional resources and energy requirements.

Within the Simultaneous Wireless Information and Power Transfer (SWIPT) paradigm, security challenges are especially critical. SWIPT-enabled systems—particularly Battery-Free Sensing Nodes (BFSNs)—rely on far-field WPT for power supply, while maintaining data communication. However, strict energy constraints prevent the use of traditional security protocols, which are computationally and power-intensive. As a result, these devices depend on lightweight wireless technologies such as LoRaWAN with static-key Activation by Personalization (ABP) encryption, or BLE in broadcast mode, which avoids pairing and connection procedures [13], [14].

### A. Related Works

A wide range of security mechanisms have been explored in the context of SWIPT systems, spanning cooperative relaying,

This research was partially funded, in whole or in part, by the French National Research Agency (ANR) under the project "ANR-25-CE39-5853-01". (Corresponding authors: Taki E. Djidjekh, and Alexandru Takacs).

Taki E. Djidjekh is with LAAS-CNRS, Université de Toulouse, CNRS, 31400 Toulouse, France. (e-mail: taki-eddine.djidjekh@laas.fr)

Alexandru Takacs is with LAAS-CNRS, Université de Toulouse, CNRS, UPS, 31400 Toulouse, France. (e-mail: alexandru.takacs@laas.fr)

Gaël Loubet is with LAAS-CNRS, Université de Toulouse, CNRS, INSA, 31400 Toulouse, France. (e-mail: gael.loubet@laas.fr)

Lamoussa Sanogo is with LAAS-CNRS, Université de Toulouse, CNRS, 31400 Toulouse, France. (e-mail: lamoussa.sanogo@laas.fr)

Daniela Dragomirescu is with LAAS-CNRS, Université de Toulouse, CNRS, INSA, 31400 Toulouse, France. (e-mail: daniela.dragomirescu@laas.fr).

jamming strategies, intelligent reflecting surfaces, protocol-level solutions, and physical-layer techniques, including the approach that is further expanded in this article. Table I summarizes representative approaches, highlighting their core methods, main security and energy benefits, and the limitations that affect their scalability and suitability for battery-free Wireless Sensor Networks.

TABLE I
COMPARATIVE OVERVIEW OF SWIPT-BASED SECURITY APPROACHES

| Year | Reference | Approach | Key Features | Advantages | Limitations |
|---|---|---|---|---|---|
| 2020 | [15] | Full-duplex self-jamming with power splitting | Receiver harvests and transmits artificial noise while decoding. | Secures *via* self-jamming; jamming power comes from harvested energy. | Complex self-interference control; evaluated only through simulation models. |
| 2021 | [16] | Unmanned Aerial Vehicle (UAV) directional modulation (millimeter-wave) | Angle and range selective beams; power split and trajectory optimization algorithm. | Secures by spatial focusing (no artificial noise); power splitting supplies energy while receiving. | Unmanned Aerial Vehicle & large array dependent; complex control and coordination; evaluated only through simulation models. |
| 2021 | [17] | Robust Beamforming & Artificial Noise in Het-Networks | Joint beamforming and noise injection under imperfect Channel State Information (CSI). | Degrades eavesdroppers while improving total energy efficiency; robust to channel errors. | Complex optimization algorithms; architecture-dependent not scalable for all IoT standards; evaluated only through simulation models. |
| 2022 | [18] | Intelligent reflecting surface–assisted SWIPT | Optimize transmitter power, power-splitting, and reflector phases to strengthen the user link and weaken the eavesdropper. | Secures by shaping the channel; passive surface is energy-efficient; benefits grow with more reflecting elements. | Requires accurate channel knowledge and many controlled elements; not protocol-agnostic in practice. |
| 2022 | [19] | Millimeter-wave IoT secrecy with hybrid precoding and artificial noise | Joint digital/analog precoder, artificial noise, power split; nonlinear harvesting. | Secures by injecting artificial noise while beamforming toward the user; reduce hardware energy. | Millimeter-wave (cellular) hardware dependent. Not scalable to all lightweight SWIPT IoT. |
| 2022 | [20] | Relay-assisted cognitive radio with non-orthogonal multiple access and SWIPT | Jointly optimizes user transmit powers, relay beamforming, and the power-splitting factor to maximize the secrecy rate against eavesdroppers. | Secures by shaping the physical channel; proposed path-following algorithm achieves strong secrecy with moderate computation. | Requires channel knowledge and relay infrastructure; depends on cognitive-radio and non-orthogonal multiple-access assumptions. |
| 2022 | [21] | Direct Sequence Spread Spectrum (DSSS) & SWIPT | Distinct pseudo-noise codes to label data, power, and key delivery; send encryption keys within spread power transfer; encrypt uplink data. | Secures by hiding data and using symmetric encryption with lower energy. | Hardware complexity; only compatible with DSSS-based standards; results rely on modeled assumptions. |
| 2023 | [22] | Cooperative amplify-and-forward relay for SWIPT sensor networks | Best-source selection; relay uses power splitting to optimize outage and intercept probability. | Link diversity and source selection lower intercept probability; tuning the power split controls intercept risk. | Requires an energy harvesting relay; assumes channel knowledge; important coordination protocol overhead. |
| 2023 | [23] | Two-way SWIPT relay with friendly jammer | Joint friendly jammer and geometric programming to maximize secrecy with time switching ratio of relay and transmit power. | Jamming and time switching increase secrecy capacity. | Requires interference management; imperfect cancellation can hurt user signals. |
| 2024 | [24] | Deep-learning beamforming for secure Internet of Things SWIPT | Base station uses a deep-learning to select information and energy beams that satisfy a secrecy target while maximizing harvested energy. | Improves secrecy and harvested energy jointly; no changes to device-side communication protocols. | Needs central training, channel and energy measurements, and multi-antenna hardware. |
| 2024 | [25] | Graph Recurrent Neural Network (GRNN) eavesdropping detection for SWIPT | GRNN model of network characteristics to accurately detect the eavesdropping behaviors in the network. | Centralized at base station; efficient detection of eavesdroppers by their attacks. | Needs network feature collection (CSI/RSSI/energy) and model training; not adaptable for silent eavesdroppers; detection, not prevention. |
| 2024 | [26], [27] | Proposed Backscattering Security Concept | Adding authentication to the power link using backscattering with minimal energy; data link kept untouched. | Near-zero additional energy at the constrained sensing node level; protocol agnostic; real world proof of concept. | Needs optimization to increase backscattering range and collision handling |

In summary, although existing SWIPT-based security schemes improve secrecy and energy efficiency under different conditions, most rely on additional hardware (e.g., relays, friendly jammers, intelligent reflecting surfaces, unmanned aerial vehicles) that increases cost, synchronization burdens, and points of failure; or on complex processing (convex optimization, deep-learning control, multi-antenna beamforming) that assumes accurate, continuous channel knowledge which can be unpractical in dynamic networks. Most are also protocol specific (e.g., DSSS, NOMA), limiting their applicability to other IoT stacks. Moreover, the majority of these works are algorithmic and validated only by analysis/simulation, with limited real-world testbed evidence, leaving deployment risks underexplored. By contrast, the work expanded in this paper (highlighted in Table I) is protocol-agnostic and energy-efficient, making it more suitable for ultra-low-power SWIPT IoT devices, particularly battery-free wireless sensors.

*B. Contributions*

This article addresses theses SWIPT-based security gaps by expanding on a promising solution originally presented in [26], [27]. Specifically designed for IoT applications, this protocol-agnostic approach introduces a lightweight, energy-efficient security mechanism grounded in backscatter communication and wireless power transfer. In addition to positioning this solution within a comparative analysis of related works, the main contributions of this paper are:

- *Innovative, Protocol-Agnostic Security Mechanism:* A continuation and in-depth exploration of the backscatter-based security layer from [26] and [27], which secures SWIPT-enabled IoT devices with minimal increase in energy consumption and hardware. This article elaborates on its architecture, operation, and scalability for broader deployment.
- *Conceptual Security Framework:* A set of lightweight, backscatter-driven authentication strategies that operate without activating RF transceivers, enabling secure identification in IoT devices within the SWIPT concept.
- *Security Testing:* Emulation of replay attacks to evaluate vulnerabilities in standard LoRaWAN encryption and confirm the relevance of the proposed identification mechanism.
- *Experimental Validation:* Implementation and testing using LoRaWAN-compatible, battery-free sensing nodes within a real Wireless Sensor Network. This confirms the proposed solution's practicality and robustness under real-world SWIPT operation scenario.
- *Compact Antennas for Practical Applications:* Demonstrations using low-profile, low-gain antennas to validate the suitability of the solution for space-constrained IoT deployments.

The remainder of this paper is organized as follows: Section II details the proposed WPT backscattering-based security concept and proposes multiple implementation strategies. Section III presents experimental validation, including replay attack emulation and performance evaluation. Section IV discusses the results, limitations, and potential areas for future improvements. Finally, Section V concludes the paper by summarizing the key contributions and their implications for IoT deployments.

## II. WPT Backscatter-Based Security and Implementation Strategies For Enhancing IoT Security

Within the framework of SWIPT for IoT applications, a security and identification mechanism based on WPT offers a promising means to significantly enhance system resilience against cyber-attacks while introducing minimal additional power consumption. This section provides a detailed exploration of this backscatter-based security approach and presents various implementation strategies aimed at maximizing its effectiveness, scalability, and integration within existing IoT infrastructures.

*A. WPT Backscatter-based Security*

In a typical SWIPT configuration, a Communicating Node (CN) drives a dedicated RF source operating in the ISM radio band to transmit a power waveform (P-wave) in the downlink direction to multiple BFSNs within a WSN. These BFSNs harvest energy from the received P-wave to power their operations, and employ separate data waveforms based on their standard network protocols—such as LoRaWAN, Bluetooth Low Energy, or ZigBee—to transmit sensor data back to the CN [13], [14].

Building on this architecture, the proposed concept introduces an additional security layer that operates independently of the communication protocol. Specifically, BFSNs can transmit an identification information back to the CN by modulating a backscattered P-wave with a Private Key (PvK), effectively embedding an uplink signal within the original power waveform. The electromagnetic fingerprint of the original P-wave—defined by parameters such as carrier frequency, modulation, and power spectral density—serves as a Public Key (PK), implemented at the physical layer. This electromagnetic fingerprint can be dynamically varied in real time using a digital key, as long as the modifications remain compliant with the regulations of the selected ISM band. This backscatter-based communication enables secure identification without activating the RF transmitters of the BFSNs. Fig.1 illustrates both the standard SWIPT architecture and the proposed enhanced concept, highlighting the hardware architectures of the Sensing and Communicating Nodes, as well as the structure of the electromagnetic waveforms exchanged between them.

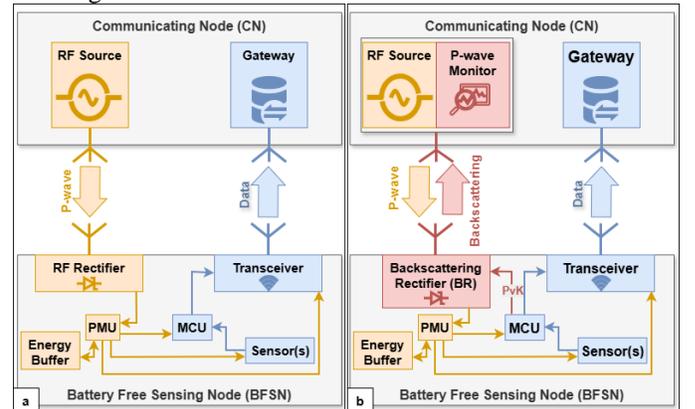

**Fig. 1.** (a) Standard SWIPT architecture; (b) Proposed backscatter-based security in SWIPT architecture. Both panels show node configurations and waveform exchanges.

At the hardware level, the core innovation lies in replacing the conventional RF rectifier within the BFSNs with a Backscattering Rectifier (BR) [27] and incorporating a dedicated Power-Wave Monitor at the CN. The BR is controlled by a single General-Purpose Input/Output (GPIO) pin from the BFSN's microcontroller unit (MCU) and operates in two distinct modes. In harvesting mode, the BR is impedance-matched to the antenna, enabling efficient conversion of the incoming RF P-wave into DC energy, which is routed to the Power Management Unit (PMU) for storage. In backscattering mode, the BR is deliberately impedance-mismatched from the antenna, causing the P-wave to be reflected and modulated. This modulated reflection serves as a lightweight, energy-efficient identification uplink, which can be detected and decoded by the Power-Wave Monitor at the CN, adding a secure identification layer to the SWIPT architecture without requiring active RF transmission.

In summary, this approach leverages backscatter modulation of the RF P-wave to embed secure identification within a WPT uplink channel. By digitally encoding a unique PvK through the BFSN's BR, the system enables real-time authentication at the CN by comparing the backscattered signal against the known P-wave serving as the Public Key. Following successful identification, the BFSNs transmit their sensor data over the standard IoT protocols. This ensures robust security and reliable data communication without altering existing network protocols or incurring significant additional energy consumption.

### B. Implementation Strategies for WPT Backscatter-Based Identification

This section proposes practical strategies for implementing the backscatter-based identification mechanism introduced in the SWIPT security framework. Each strategy aims to balance minimal energy consumption with robust identification, making them suitable for integration into BFSNs.

*1) Private Key-Based Backscattering*

This strategy enhances security in SWIPT-based systems by modulating the backscattered signal using a unique digital PvK. The key can be selected from a predefined code table or generated dynamically using a lightweight encryption algorithm such as the Advanced Encryption Standard (AES). Owing to its low computational and energy requirements, AES is well-suited for implementation on the microcontrollers of BFSNs and the Power-Wave Monitor within the Communicating Node. By embedding the PvK within the backscattered signal, this approach significantly increases resistance to spoofing and unauthorized replication. This concept was experimentally validated using On-Off Keying (OOK) amplitude modulation.

*2) Public Key Frequency-Hopping*

Here the system employs a frequency-hopping strategy by dynamically varying the frequency and thus the electromagnetic fingerprint of the P-wave generated by the RF source at CN level. Both RF source and the Backscattering Rectifier supports operation across a wide RF input band (863–870 MHz, European ISM band), allowing the CN to transmit the P-wave—serving as the Public Key—at a randomly selected frequency for each authentication event. The BFSN then reflects the signal at this specific frequency, enabling the CN to verify both the timing and spectral accuracy of the backscattered response. This per-frame frequency agility prevents adversaries from successfully replaying or emulating previously captured identification signals, thereby reinforcing the security of the WPT-based identification layer.

*3) Dual-Key encoding mechanism*

This strategy strengthens security by embedding a digital Public Key into the WPT waveform (P-wave) generated by the RF source at the CN, as illustrated in Fig. 1. The BFSN dynamically backscatters the incoming P-wave while simultaneously encoding the PvK using the BR, which is controlled by the BFSN's microcontroller. The CN's P-wave Monitor then correlates the real-time backscattered response with the expected Public and Private Key combination to validate the authenticity of the BFSN. This approach effectively mitigates replay and spoofing attacks without significantly increasing computational complexity or power consumption at the sensing node side.

Fig. 2 illustrates the system architecture implementing the dual-key encoding strategy, highlighting how the two digital public and private keys are jointly embedded in the backscattered signal for secure identification.

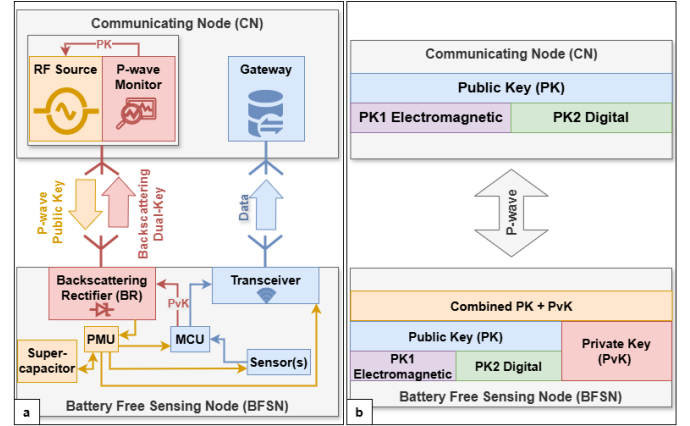

**Fig. 2.** (a) SWIPT Architecture with the Dual-Key WPT Backscattering-based Security Layer; (b) Dual-Key Security Architecture.

### III. EXPERIMENTAL VALIDATION

In this section, we present the experimental validation of the proposed solution, beginning with the emulation of a replay attack targeting a BFSN to demonstrate existing vulnerabilities in conventional setups. We then integrate the proposed Backscattering Rectifier into the same BFSN to evaluate the effectiveness of the authentication mechanism. This validation includes characterization of the BR's behavior and the deployment of the complete setup within a WSN environment.

### A. Replay Attack Emulation

To evaluate the vulnerabilities of conventional LoRaWAN-based BFSNs, we first implemented a replay attack on a BFSN developed in [13] for SHM applications, which uses ABP encryption due to its constrained power and computing resources. ABP relies on static keys and bypasses the join procedure, making it particularly vulnerable to replay attacks [28].

Replay attacks were carried out using two attacker setups based on: (i) a HackRF Software-Defined Radio (SDR) and (ii) a Microchip RN2483 LoRa transceiver. The experimental setup, shown in Fig. 3, includes a LoRaWAN gateway, an RF power source, the BFSN, and the attacker device. While the components were placed in close proximity for testing, the inherent long-range nature of LoRaWAN allows similar attacks to be executed from much greater distances.

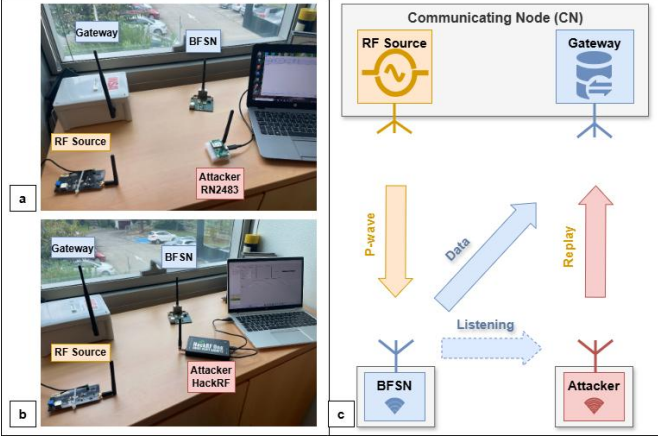

**Fig. 3.** Experimental setup for replay attack: (a) setup with RN2483 LoRa transceiver as attacker device, (b) setup with HackRF SDR as attacker device, (c) diagram of replay attack process.

A replay attack using the RN2483 transceiver, as presented in Fig. 3 (a), was implemented as follows: (i) the RN2483 was configured to receive on the BFSN's uplink frequency, and an encrypted message sent by the BFSN was intercepted and stored, (ii) the unencrypted frame counter, which remains static under ABP encryption, was extracted, (iii) the RN2483 was then switched to transmit mode and used to replay the same message on a different LoRaWAN frequency channel, (iv) since LoRaWAN uses a channel-hopping mechanism, replaying the message on another valid channel aligned with expected network behavior, (v) the LoRaWAN network server accepted the replayed frame without detection, as shown in Fig. 4.

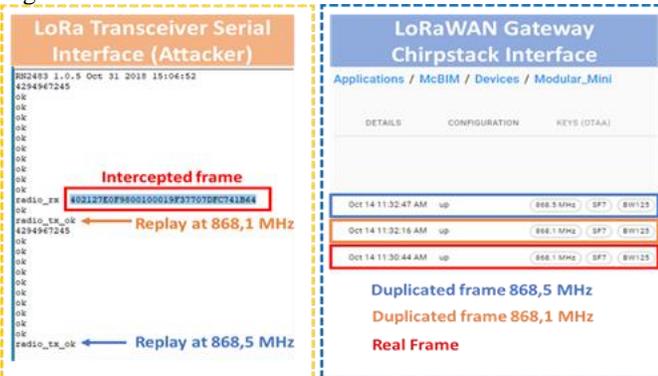

**Fig. 4.** LoRa transceiver serial interface displaying captured frame (left) and gateway ChirpStack network application with original and duplicated frames (right).

Replay Attack using the HackRF SDR, as presented in Fig. 3 (b), was carried out similarly: (i) the HackRF was set to passively monitor the BFSN's LoRaWAN uplink frequency and a valid frame was captured and recorded, (ii) the recorded message was then replayed on the same frequency, (iii) the LoRaWAN network server accepted the replayed frame, as ABP does not verify frame origin beyond static key matching.

These results highlight a critical security weakness: the limited computational and energy resources of BFSNs often necessitates lightweight encryption like LoRaWAN ABP, leaving them exposed to basic replay attacks and denial-of-service (DoS). In practice, an adversary can repeatedly replay valid-looking frames to consume airtime and gateway processing or inject continuous transmissions to jam the uplink, thereby degrading availability. This underscores the urgent need for robust, low-power security mechanisms tailored to resource-constrained IoT devices. Additional LoRaWAN vulnerabilities discussed in [28] further emphasize the necessity for more advanced protection strategies.

*B. Characterization of the Backscattering Rectifier*

The Backscattering Rectifier comprises a compact RF rectifier circuit that includes an LC impedance-matching network, a diode pair (SMS7630-005LF from Skyworks Solutions, Inc.) integrated into a single package, and an N-channel MOSFET (BSS123 from ON Semiconductor) configured in a diode-connected arrangement, as illustrated in Fig. 5. Under normal conditions with the gate-to-source voltage ($V_{GS}$) at 0 V, the BR achieves optimal impedance matching with the antenna port, enabling efficient energy harvesting. When activated by the microcontroller unit (MCU) via a GPIO pin, the MOSFET's effective ON-resistance is altered, disrupting this impedance match and reflecting the incoming RF power wave. By dynamically modulating $V_{GS}$ via a PvK, the circuit generates a controlled, modulated backscatter signal. This approach facilitates highly efficient continuous energy harvesting, with only brief intervals allocated for identification through backscattering, thereby minimizing power consumption while enhancing security.

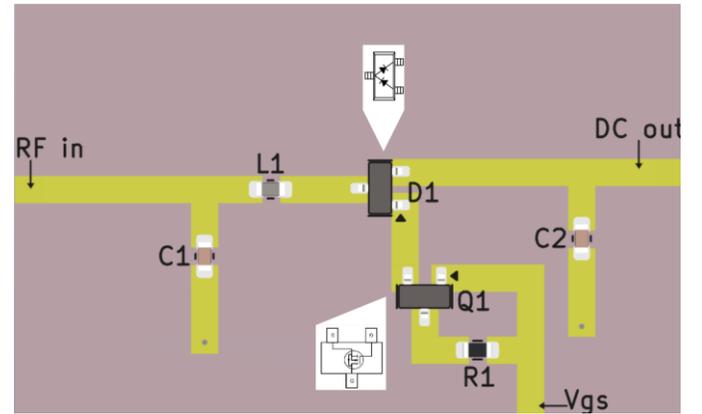

**Fig. 5**. Backscattering Rectifier circuit layout.

The BR was designed to operate in the 868 MHz ISM band, chosen as a compromise between regulation limits in term of equivalent radiated power, antenna size and effective WPT range.

Initial design validation was conducted through simulations in Advanced Design System (ADS) software. The circuit was then fabricated on a two-layer PCB using a 0.8 mm FR4 substrate. Experimental characterization focused on three key performance metrics: the reflection coefficient ($S_{11}$) to assess impedance matching, the RF-to-DC conversion efficiency across various input power levels, and the backscattering capability of the BR, specifically its ability to generate a modulated backscattered signal for identification purposes.

*1) Reflection coefficient ($S_{11}$)*

The reflection coefficient was measured at the BR RF input using a Vector Network Analyzer (Rohde & Schwarz ZNL3) to evaluate impedance matching across the 868 MHz ISM band. Fig. 6 (a) presents the measured $S_{11}$ values for various input power levels, illustrating the rectifier's matching performance under typical operating conditions when $V_{GS}$ set to 0V.

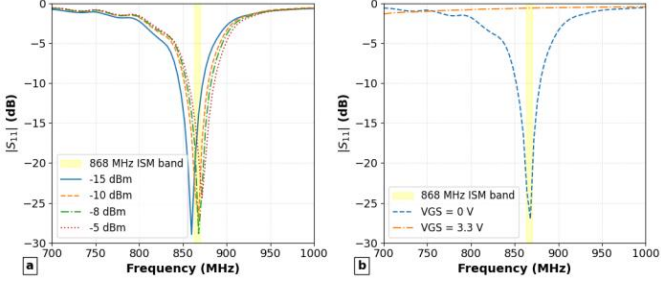

**Fig. 6**. Measured $S_{11}$ magnitude of the Backscattering Rectifier: (a) at various RF input power levels with $V_{GS} = 0$ V; (b) at –10 dBm with $V_{GS} = 0$ V and 3.3 V. Shaded area indicates the 868 MHz ISM band.

The measured $S_{11}$ curves at typical input power levels remain below -10 dB within the shaded 868 MHz ISM band. Despite variations in impedance due to the nonlinearity of the Schottky diode, the rectifier maintains effective matching across this frequency range.

The $S_{11}$ of the BR was also measured with $V_{GS} = 3.3$ V to evaluate its backscattering behavior. As shown in Fig. 6 (b), the BR exhibits strong reflection in this mode, with $S_{11} \geq -0.6$ dB, in contrast to the well-matched condition observed when $V_{GS} = 0$ V, confirming the circuit's ability to toggle between harvesting and reflective states.

*2) RF-to-DC conversion efficiency*

The RF-to-DC conversion efficiency was evaluated by applying controlled RF input power to the BR using an Anritsu MG3694B RF signal generator and measuring the DC output voltage across a 10 kΩ load resistor with a Keithley 2000 precision multimeter. Both instruments were controlled through a custom LabVIEW application developed specifically for rectifier characterization. Measurements were performed with $V_{GS}$ set to 0 V, placing the BR in energy harvesting mode. As illustrated in Fig. 7, the rectifier achieves an RF-to-DC conversion efficiency exceeding 20 % across a broad range of typical input power levels within the 868 MHz ISM frequency band. Notably, the BR delivers approximately 15 µW DC power at an RF input power of –13 dBm, sufficient for the cold-start operation of the Power Management Unit (BQ25504, Texas Instruments [25]) used in the BFSN used in the replay attack emulation.

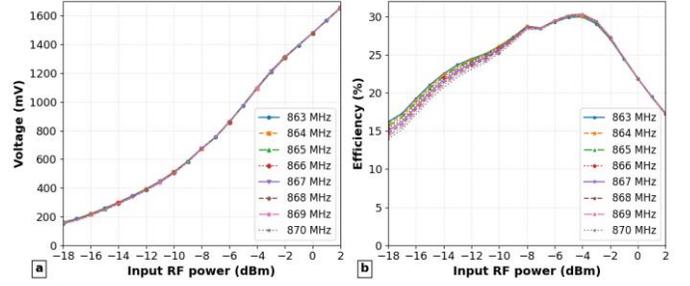

**Fig. 7**. (a) DC output voltage; (b) RF-to-DC conversion efficiency of the Backscattering Rectifier vs. RF input power. Traces represent different frequencies within the 868 MHz ISM band.

*3) Backscattering capability*

To further assess the backscattering capability of the BR, a wired experimental setup was implemented, as shown in Fig 8. The setup consisted of an Anritsu MG3694B RF signal generator generating a continuous 868 MHz P-wave, a Keithley 3390 waveform generator supplying a 0–3.3 V square wave to control $V_{GS}$, an Aerotek C11-1FFF RF circulator, and a Tektronix RSA306B USB spectrum analyzer acting as the P-wave monitor. The square wave toggled the BR between harvesting and backscattering modes, modulating the reflection of the incoming RF signal. This modulated backscattered signal was routed back through the circulator and captured by the spectrum analyzer. The results showed stable modulation with a clear dynamic range—defined as the difference in reflected power between the high and low states—across modulation frequencies up to 100 kHz. The backscattered signal exhibited a dynamic range of approximately 20 dB, as illustrated in Fig. 9.

This measured dynamic range can be analytically roughly estimated by considering the known setup parameters. With an RF input power ($P_{RF}$) delivered by RF source of –10 dBm, circulator isolation (I) of minimum 20 dB, and forward insertion losses ($L_f$) around 1 dB (including circulator and cable losses), the leakage baseline at the spectrum analyzer is:

$$P_{leak}[dBm] = P_{RF} - I = -10\ dBm - 20\ dB = -30\ dB \quad (1)$$

At –10 dBm RF input, the measured reflection coefficient ($S_{11}$) of BR was approximately –0.6 dB for $V_{GS} = 3.3$ V. The reflected power ($P_{refl}$) at the BR is thus expressed as:

$$P_{refl}[dBm] = P_{RF} - 2L_f + S_{11} = -10\ dBm - 1.6\ dB - 0.6\ dB = -12.2\ dBm \quad (2)$$

Considering both leakage and reflected components, the dynamic range (ΔP) estimated at the spectrum analyzer operating as P-wave monitor input is approximately:

$$\Delta P = P_{refl} - P_{leak} = -12.2\ dBm + 30\ dBm = 17.8\ dB \quad (3)$$

This analytical value is in close agreement with the measured 16.6 dB dynamic range shown in Fig. 9, confirming that the Backscattering Rectifier effectively modulates the reflected signal under the given test conditions.

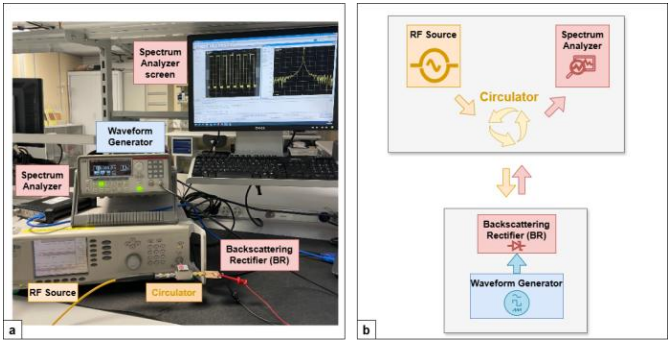

**Fig. 8.** Wired experimental setup to evaluate the backscattering capability of the backscattering rectifier: (a) Photo of the setup; (b) Schematic diagram.

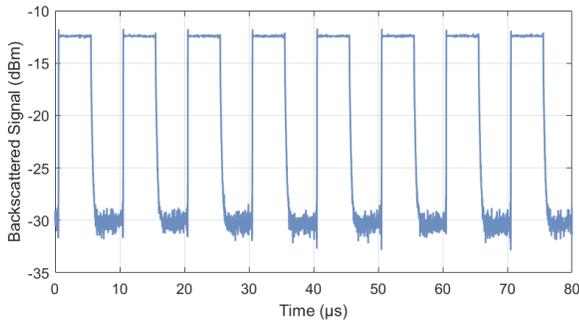

**Fig. 9.** Backscattered signal captured by the spectrum analyzer (P-wave monitor) at a modulation frequency of 100 kHz, with a -10 dBm continuous P-wave input at 868 MHz.

*C. Experimental Validation of Backscatter-Based Security*

*1) BR integration into LoRaWAN BFSN and the measurement of the power consumption during a typical operation cycle*

The BR was integrated into the BFSN [13], as depicted in Fig. 10. The original RF rectifier was disabled, and the BR output was connected to the DC input of the PMU. Additionally, a GPIO pin from the BFSN's MCU was directly connected to the BR's gate ($V_{GS}$) to enable mode switching between harvesting and backscattering.

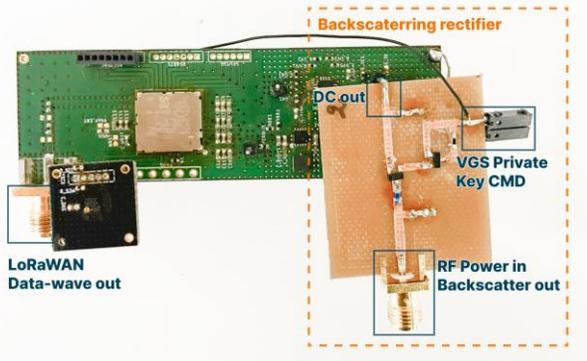

**Fig. 10.** Backscattering Rectifier integrated into a Battery-Free LoRaWAN Sensing Node.

The BFSN was reprogrammed to perform a brief 2 ms identification sequence transmitting a 16-byte PvK, just before initiating the LoRaWAN ABP transmission. This PvK was Manchester-encoded in real time by toggling the GPIO pin controlling the BR's MOSFET gate. During this short backscattering period, the MCU remain active, but the RF transceiver stay in sleep mode to minimize overall power consumption.

Fig. 11 illustrates the measured timing and power-consumption profile of the BFSN's full operational cycle. The highest power consumption is observed during initialization and LoRaWAN transmission phases. In contrast, the 2 ms backscattering sequence contributes negligibly to the total energy usage, confirming the energy-efficient nature of the proposed identification mechanism.

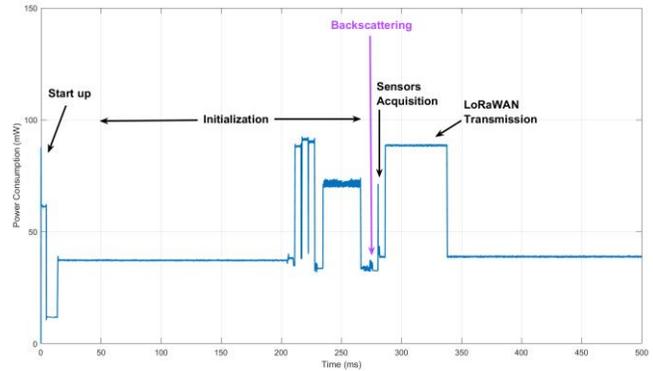

**Fig. 11.** Power consumption profile of the LoRaWAN BFSN operation cycle with the authentication backscattering concept

*2) Single-Node Wireless Validation*

Initial wireless tests were conducted using the LoRaWAN BFSN in a real-world environment outside the anechoic chamber to validate the proposed security concept under practical conditions. Two identical patch antennas (+9.2 dBi gain) were used for the WPT link (one for the BFSN's BR and the other for the P-wave Monitor). The wireless setup mirrored the wired configuration including an Anritsu MG3694B RF signal generator, an Aerotek C11-1FFF RF circulator providing at least 20 dB isolation, and a Tektronix RSA306B USB Spectrum Analyzer functioning as the P-wave Monitor. For the LoRaWAN data link, standard +2.5 dBi monopole antennas were used to establish communication between the BFSN and the gateway. Fig. 12 shows both the schematic and a photograph of the wireless test setup.

The distance between the BFSN and the Communicating Node (P-wave Monitor) was set to 1.61 meters. It's important to note that in this WSN application, the LoRaWAN protocol was selected primarily for its low power consumption and its balanced compromise between antenna size and effective communication through concrete structures, as required in SHM applications [13], rather than for its long-range capability. Indeed, in SWIPT applications, the communication range is inherently limited by the WPT link rather than the data link. The RF source emitted a continuous 868 MHz P-wave. With +15 dBm at the RF port and the +9.2 dBi patch antenna, the effective isotropic radiated power was about +24.2 dBm EIRP, supplying energy to charge the BFSN's super-capacitor. Once sufficiently charged, the BFSN transmitted its identification sequence by backscattering the P-wave. This backscattered identification signal was captured by the Spectrum Analyzer acting as the P-wave Monitor, as illustrated in Fig. 13. After successful identification via the WPT link, the BFSN transmitted temperature and relative humidity measurements using LoRaWAN with ABP encryption. These sensor data were subsequently received by the LoRaWAN gateway and retrieved through the ChirpStack network server, confirming that data acceptance or rejection can be effectively controlled based on prior identification via backscatter security layer.

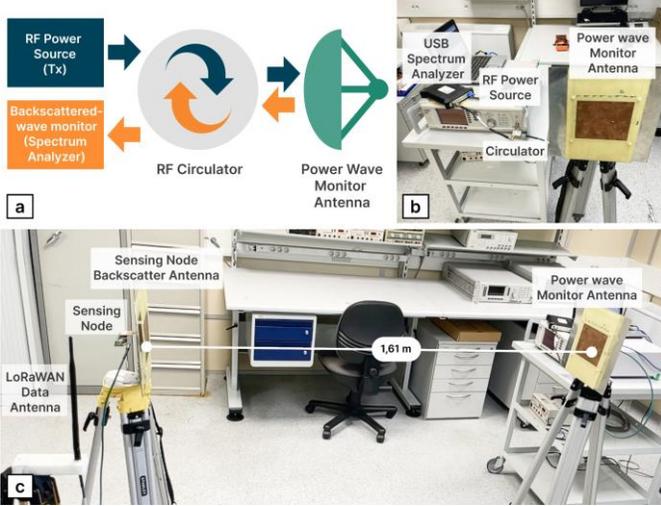

**Fig. 12.** Experimental setup: (a) schematic diagram; (b) Communicating Node; (c) Sensing Node illuminated by P-wave with backscattered signal monitored by the spectrum analyzer.

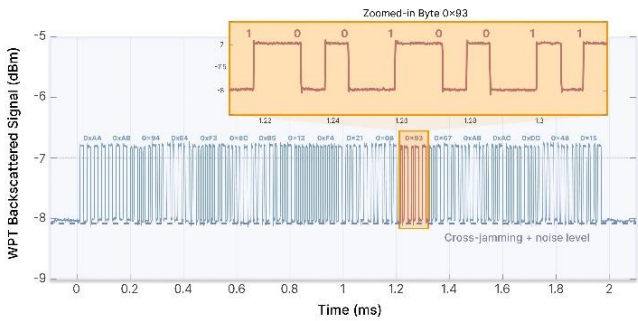

**Fig. 13.** WPT backscattered identification code signal of LoRaWAN Sensing Node retrieved by the spectrum analyzer acting as P-wave monitor.

A limited dynamic range of approximately 1 dB was observed in the wireless test, Fig. 13, (compared to 16.6 dB in the wired setup Fig. 9). This dynamic range limitation is due primarily to three factors:

- **Circulator leakage (cross-jamming):** The Aerotek C11-1FFF circulator provides at least 20 dB isolation between Tx and Rx ports, with the RF source set at +15 dBm at 868MHz, this yields a substantial maximum leakage at the spectrum analyzer input:

$$P_{leak} = P_{Tx} - I \approx 15 dBm - 20\ dB \approx -5\ dBm \quad (4)$$

- **Environmental backscatter:** Reflections from walls, floors, and nearby objects in a real-world environment further increase the background noise level, obscuring weak signals (Fig. 12(c)).

- **Limited power return from the BR:** Two +9.2 dBi patch antennas, separated by 1.61 m, were used. Considering the free-space path loss at 868 MHz, the forward path attenuation is approximately:

$$FSPL = 20 Log_{10}\left(\frac{4\pi d}{\lambda}\right) \approx 35.4\ dB \quad (5)$$

Even though the BR receives adequate power to operate, the backscattered signal experiences the same substantial losses on the return path, resulting in a weak signal at the input of the spectrum analyzer. Combined with the dominant circulator leakage, this significantly reduces the measurable dynamic range.

*3) Multi-node experimentation with compact antenna*

To experiment the generality and scalability of the BR security concept, a minimalist WSN comprising two LoRaWAN BFSNs and one Communicating Node (integrating a RF source and a spectrum analyzer operating as P-wave monitor) was implemented. A second BFSN with a different form factor, also originating from the SHM application described in [13], was integrated. The BR was connected similarly to the first node, and the embedded software was modified to generate the Manchester-coded identification sequence prior to the LoRaWAN transmission. Fig. 14 illustrates this second BFSN and the complete two-node setup.

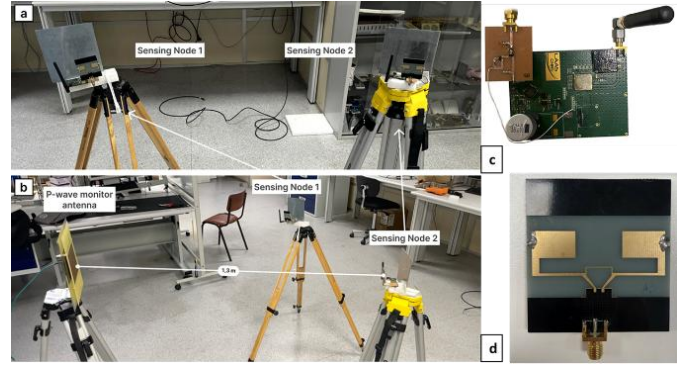

**Fig. 14.** (a) Photograph of the two BFSNs; (b) complete multi-node experimental setup; (c) close-up of the second BFSN; (d) compact dipole antenna used for the WPT link.

As shown in Fig. 14(d), the bulky patch antennas on the WPT link were replaced by compact in-house dipoles (gain ≈ +2 dBi, 40 × 30 × 10 mm) [29]. Two BFSNs were positioned 1.3 m from a single RF source radiating +20 dBm CW at 868 MHz. The CN retained its +9.2 dBi patch antenna, while standard 2 dBi monopoles were kept for the LoRaWAN data link (Fig. 14c).

Fig. 15 shows the backscattered signal captured from the second BFSN. Both nodes exhibited a reduced dynamic range of approximately 0.5 dB, primarily due to increased cross-jamming, environmental noise in the wireless test environment, and the lower gain of the compact dipole antennas. Despite these limitations, the identification sequences from both BFSNs were reliably captured by the Spectrum Analyzer, and uplink sensor data were successfully received by the LoRaWAN network server. Additional frequency-sweep tests across the 868 MHz ISM band confirmed consistent dynamic-range performance, though potential collisions between backscattered signals remain a constraint in simultaneous multi-node scenarios.

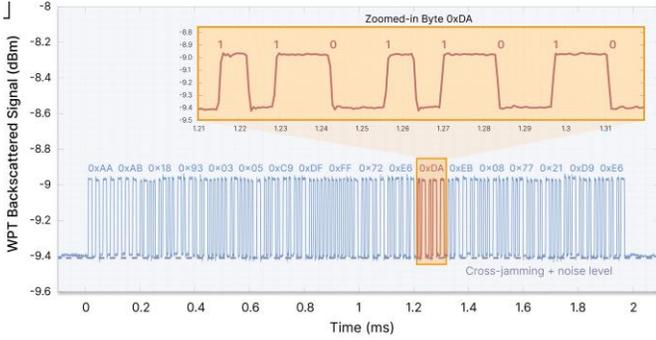

**Fig. 15.** Backscattered frame captured with compact WPT antenna from the second Sensing Node.

## IV. Discussion

The implementation of a WPT-based security and identification solution in a SWIPT context for IoT applications demonstrates a promising enhancement in security without imposing significant computational burdens on BFSNs. The experiments validated the feasibility of using Backscattering Rectifiers to modulate identification signals within a Wireless Sensor Network. This section elaborates on how the concept prevents replay attacks, key outcomes, advantages, limitations, and areas for future improvements, with an emphasis on addressing scalability and environmental challenges.

### A. Replay Attack Prevention

The experiments confirmed that transmitting a backscattered Private Key over the WPT link adds a valuable security layer for BFSN identification. As shown in Fig. 16, the P-wave Monitor can successfully detect and reject unauthorized or replayed signals by analyzing the backscattered uplink component of the P-wave.

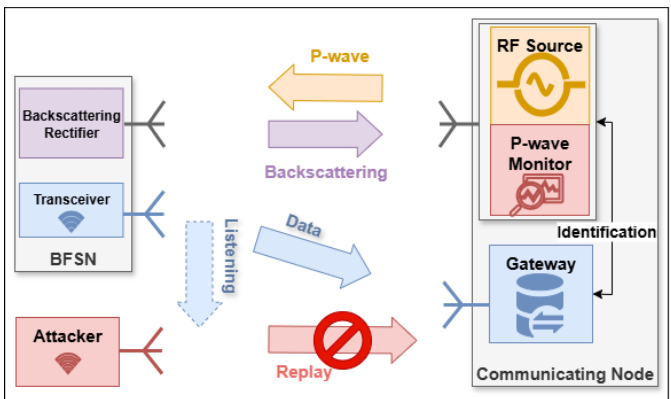

**Fig. 16.** Schematic diagram of replay attack countering with WPT-based security framework.

However, the system remains vulnerable to advanced replay attacks where an attacker could capture and replay the PvK waveform to impersonate a legitimate device. To address this, strategies such as Public Key frequency-hopping and dual-key modulation (described in Section II-B) introduce randomness and real-time signal variation. These methods significantly increase resistance to replay attempts by requiring precise timing and frequency synchronization, thus enhancing security and robustness in noisy or interference-prone environments.

### B. Experimental Outcomes

**Enhanced Security:** By incorporating backscattering techniques with a minimal impact in term of energy consumption and hardware complexity on device and system level, the proposed solution adds an extra layer of security that is independent and complementary of traditional IoT communication protocols. This approach helps prevent unauthorized access and data corruption, providing a robust defense against potential replay and relay attacks [1], [5].

**Low Computational Overhead:** The Backscattering Rectifier is controlled by a single GPIO pin of the Sensing Node's MCU, requiring minimal additional processing power. This design ensures that the energy efficiency of the Sensing Node is preserved, which is crucial for battery-free or energy-constrained IoT devices.

**Scalability and Feasibility:** Experimental validation with two LoRaWAN battery-free Sensing Nodes demonstrated the successful integration of the Backscattering Rectifier, highlighting its applicability to different nodes. However, since the transmissions from these nodes were not simultaneous, the P-wave monitor's ability to handle multiple WPT uplinks concurrently was not tested.

**Environmental Adaptability:** The experiments were conducted in a laboratory setting outside an anechoic chamber, where the backscattered fingerprint of the room, due to measurement instruments, antennas, and other objects, contributed to increased cross-jamming. While the system demonstrated effective functionality under these conditions, further testing in environments with higher levels of environmental noise is needed to better assess and potentially improve its reliability in real-world deployments.

### C. Advantages of the WPT-based Identification Solution

**Protocol Independence:** The WPT-based identification layer operates independently of the underlying communication protocol, making it highly adaptable and easy to integrate into BFSNs using various protocols, such as LoRaWAN, BLE, or Zigbee. In [30] the same concepts was implemented in a BLE based BFSN. This flexibility ensures broad compatibility across diverse IoT networks.

**Energy and Computational Efficiency:** The backscattering of the WPT P-wave requires minimal energy, making it particularly suitable for energy-constrained devices. Moreover, the Backscattering Rectifier is controlled by a single GPIO pin, which demands negligible power, and generating a Private Key involves minimal computational resources, ensuring low energy consumption without compromising functionality.

**Scalability and Versatility:** The system's lightweight design and protocol independence make it easily scalable to multiple nodes within a WSN. The use of license-free ISM frequency bands for backscattering and energy harvesting further enhances deployment flexibility, enabling its use in applications such as structural health monitoring, healthcare, and smart building, transport and cities.

## D. Limitations and Challenges

**Dynamic Range Constraints:** The dynamic range of the backscattered signals was constrained by cross-jamming, environmental noise, and clutter effects. Cross-jamming occurred due to interference between the backscattered signals and the primary WPT signal, while environmental noise, and unintended RF emissions, further degraded signal clarity. Clutter from surrounding equipment, antennas, and measurement instruments in the testing environment added complexity, reducing the reliability of identification signals.

**Complexity of P-wave Monitor:** The PK and PvK were implemented by amplitude modulation, thus the P-wave monitor can be easily implemented as power/amplitude detector. Implementing more complex frequency-hopping techniques and modulation could increase the complexity of the P-wave monitor at the Communicating Nodes.

## V. Future Improvements and Research Directions

### A. Minimizing Cross-jamming

Cross-jamming, caused by weakness of the backscattered signal, can be mitigated using dual-polarization backscattering rectifiers/rectenna or harmonic transponder techniques. Dual-polarization rectennas separate the P-wave and backscattering across different polarizations, while harmonic transponders operate these functions at distinct harmonic frequencies. Both methods significantly enhance signal clarity and reliability [31], [32].

### B. Improving Circulator Isolation

Employing high-isolation RF circulators tailored for the P-wave monitor effectively reduces signal interference.

### C. Antenna Optimization

Development of high-gain, compact, directive antennas for sensing nodes and RF sources improves dynamic range and signal reliability, crucial for integration into compact sensing nodes.

### D. Key Performance and Security Metrics

Experimental validation primarily focused on demonstrating low-power backscattering feasibility. In the current implementation, the backscattering dynamic range, measured using a spectrum analyzer as the P-wave monitor, was on the order of 0.5–1 dB. Future evaluations should incorporate additional quantitative metrics, including bit error rate (BER), detection reliability, robustness to replay attacks, and sensitivity to environmental fingerprinting and timing synchronization. The accurate quantification of these metrics depends on the P-wave monitor architecture (e.g., envelope detection, IQ demodulation, or RCS-based processing) and is therefore identified as an important direction for future work.

### E. Integration with Packet-Based Payload Protection

Future research will also explore the integration of WPT-based physical-layer identification with lightweight cryptographic payload protection, such as AES-based schemes, in generic packet-based communication systems. In this context, physical-layer identification can enable early packet filtering and device authentication, reducing cryptographic overhead while enhancing protection against replay and injection attacks by contributing to the data encryption process.

## VI. Conclusion

This work demonstrates the feasibility of integrating a lightweight, protocol-agnostic security and identification layer directly into the wireless power link of battery-free sensing nodes. By embedding a private-key signature onto the backscattered component of the power waveform, the system effectively mitigates replay and denial-of-service (DoS) attacks without introducing measurable computational or energy overhead on the node's MCU.

Experiments with LoRaWAN BFSNs verified robust identification, effective replay-attack prevention, and seamless coexistence with conventional data channels. Preliminary results presented in [30] also confirmed the feasibility of implementing the same backscatter-based security concepts using the BLE protocol. The architecture remains highly extensible: enhancements such as dual-key modulation, frequency hopping, dual-polarization rectennas, harmonic-transponder rectifiers, high-isolation circulators, and optimized high-gain antennas can further improve system security robustness, operating range, reliability and scalability across diverse IoT deployments.

While the current P-wave monitor implemented by using a spectrum analyzer handles only a limited number of simultaneous uplinks, these results provide a solid foundation for real-world deployments spanning WSNs constrained applications such as structural health monitoring. Future work will focus on scaling the P-wave monitor to dense networks, rigorously characterizing performance in high-noise industrial settings, and integrating the countermeasures and implementation strategies outlined above to deliver a robust and reliable security for SWIPT WSNs.

## Acknowledgment

The authors extend their gratitude to Alassane Sidibé for designing the compact dipole antenna used in the experimental setups. They also acknowledge the assistance of Quentin Bernyer in the experimental setups.

## References

[1] A. Mosenia and N. K. Jha, "A Comprehensive Study of Security of Internet-of-Things," IEEE Trans. Emerg. Topics Comput., vol. 5, no. 4, pp. 586–602, Oct. 2017, doi: 10.1109/TETC.2016.2606384.

[2] J. Singh, G. Singh, and S. Negi, "Evaluating Security Principals and Technologies to Overcome Security Threats in IoT World," in 2023 2nd International Conference on Applied Artificial Intelligence and Computing (ICAAIC), Salem, India: IEEE, May 2023, pp. 1405–1410. doi: 10.1109/ICAAIC56838.2023.10141083.

[3] V. Palazzi et al., "Radiative Wireless Power Transfer: Where We Are and Where We Want to Go," IEEE Microwave, vol. 24, no. 2, pp. 57–79, Feb. 2023, doi: 10.1109/MMM.2022.3210145.

[4] G. Loubet et al., "Wirelessly Powered Battery-Free Sensing Nodes for Internet of Things Applications," IEEE Microwave, vol. 26, no. 7, pp. 26–46, Jul. 2025, doi: 10.1109/MMM.2024.3488593.

[5] S. Loukil, L. C. Fourati, A. Nayyar, and C. So-In, "Investigation on Security Risk of LoRaWAN: Compatibility Scenarios," IEEE Access, vol. 10, pp. 101825–101843, 2022, doi: 10.1109/ACCESS.2022.3208171.

[6] G. Loubet, E. Alata, A. Takacs, and D. Dragomirescu, "A Survey on the Security Challenges of Low-Power Wireless Communication Protocols for Communicating Concrete in Civil Engineerings," Sensors, vol. 23, no. 4, p. 1849, Feb. 2023, doi: 10.3390/s23041849.

[7] S. S. Dhanda, B. Singh, and P. Jindal, "Lightweight Cryptography: A Solution to Secure IoT," Wireless Pers Commun, vol. 112, no. 3, pp. 1947–1980, Jun. 2020, doi: 10.1007/s11277-020-07134-3.

[8] V. A. Thakor, M. A. Razzaque, and M. R. A. Khandaker, "Lightweight Cryptography Algorithms for Resource-Constrained IoT Devices: A Review, Comparison and Research Opportunities," IEEE Access, vol. 9, pp. 28177–28193, 2021, doi: 10.1109/ACCESS.2021.3052867.

[9] T. Schläpfer and A. Rüst, "Security on IoT devices with secure elements," 2019, doi: 10.21256/ZHAW-3350.

[10] M. F. Elrawy, A. I. Awad, and H. F. A. Hamed, "Intrusion detection systems for IoT-based smart environments: a survey," J Cloud Comp, vol. 7, no. 1, p. 21, Dec. 2018, doi: 10.1186/s13677-018-0123-6.

[11] A. Khraisat and A. Alazab, "A critical review of intrusion detection systems in the internet of things: techniques, deployment strategy, validation strategy, attacks, public datasets and challenges," Cybersecur, vol. 4, no. 1, p. 18, Mar. 2021, doi: 10.1186/s42400-021-00077-7.

[12] M. Safi, S. Dadkhah, F. Shoeleh, H. Mahdikhani, H. Molyneaux, and A. A. Ghorbani, "A Survey on IoT Profiling, Fingerprinting, and Identification," ACM Trans. Internet Things, vol. 3, no. 4, pp. 1–39, Nov. 2022, doi: 10.1145/3539736.

[13] G. Loubet, A. Sidibe, P. Herail, A. Takacs, and D. Dragomirescu, "Autonomous Industrial IoT for Civil Engineering Structural Health Monitoring," IEEE Internet Things J., vol. 11, no. 5, pp. 8921–8944, Mar. 2024, doi: 10.1109/JIOT.2023.3321958.

[14] A. Sidibe, G. Loubet, A. Takacs, and D. Dragomirescu, "A Multifunctional Battery-Free Bluetooth Low Energy Wireless Sensor Node Remotely Powered by Electromagnetic Wireless Power Transfer in Far-Field," Sensors, vol. 22, no. 11, p. 4054, May 2022, doi: 10.3390/s22114054.

[15] R. Ma, H. Wu, J. Ou, S. Yang, and Y. Gao, "Power Splitting-Based SWIPT Systems With Full-Duplex Jamming," IEEE Transactions on Vehicular Technology, vol. 69, no. 9, pp. 9822–9836, Sep. 2020, doi: 10.1109/TVT.2020.3002976.

[16] X. Sun, W. Yang, Y. Cai, and M. Wang, "Secure mmWave UAV-Enabled SWIPT Networks Based on Random Frequency Diverse Arrays," IEEE Internet of Things Journal, vol. 8, no. 1, pp. 528–540, Jan. 2021, doi: 10.1109/JIOT.2020.3005984.

[17] Y. Xu, H. Xie, C. Liang, and F. R. Yu, "Robust Secure Energy-Efficiency Optimization in SWIPT-Aided Heterogeneous Networks With a Nonlinear Energy-Harvesting Model," IEEE Internet of Things Journal, vol. 8, no. 19, pp. 14908–14919, Oct. 2021, doi: 10.1109/JIOT.2021.3072965.

[18] H. T. Thien, P.-V. Tuan, and I. Koo, "A Secure-Transmission Maximization Scheme for SWIPT Systems Assisted by an Intelligent Reflecting Surface and Deep Learning," IEEE Access, vol. 10, pp. 31851–31867, 2022, doi: 10.1109/ACCESS.2022.3159679.

[19] Z. Zhu et al., "Secrecy Rate Optimization in Nonlinear Energy Harvesting Model-Based mmWave IoT Systems With SWIPT," IEEE Systems Journal, vol. 16, no. 4, pp. 5939–5949, Dec. 2022, doi: 10.1109/JSYST.2022.3147889.

[20] C. Hu, Q. Li, Q. Zhang, and J. Qin, "Security Optimization for an AF MIMO Two-Way Relay-Assisted Cognitive Radio Nonorthogonal Multiple Access Networks With SWIPT," IEEE Transactions on Information Forensics and Security, vol. 17, pp. 1481–1496, 2022, doi: 10.1109/TIFS.2022.3163842.

[21] P. Gong, T. M. Chen, P. Xu, and Q. Chen, "DS-SWIPT: Secure Communication with Wireless Power Transfer for Internet of Things," Security and Communication Networks, vol. 2022, pp. 1–11, Jun. 2022, doi: 10.1155/2022/2650474.

[22] T. N. Nguyen et al., "Physical Layer Security in AF-Based Cooperative SWIPT Sensor Networks," IEEE Sensors Journal, vol. 23, no. 1, pp. 689–705, Jan. 2023, doi: 10.1109/JSEN.2022.3224128.

[23] M. Hayajneh and T. A. Gulliver, "Physical Layer Security in Two-Way SWIPT Relay Networks with Imperfect CSI and a Friendly Jammer," Entropy, vol. 25, no. 1, p. 122, Jan. 2023, doi: 10.3390/e25010122.

[24] V. Ganapathy, R. Ramachandran, and T. Ohtsuki, "Deep Learning Methods for Secure IoT SWIPT Networks," IEEE Internet of Things Journal, vol. 11, no. 11, pp. 19657–19677, Jun. 2024, doi: 10.1109/JIOT.2024.3368692.

[25] W. Jiang, J. Wang, K.-L. Hsiung, and H.-Y. Chen, "GRNN-Based Detection of Eavesdropping Attacks in SWIPT-Enabled Smart Grid Wireless Sensor Networks," IEEE Internet of Things Journal, vol. 11, no. 22, pp. 37381–37393, Nov. 2024, doi: 10.1109/JIOT.2024.3443277.

[26] T. E. Djidjekh, L. Sanogo, G. Loubet, A. Sidibé, D. Dragomirescu, and A. Takacs, "A New Security and Identification Concept for SWIPT Systems in IoT Applications," in 2024 IEEE/MTT-S International Microwave Symposium - IMS 2024, Washington, DC, USA: IEEE, Jun. 2024, pp. 110–113. doi: 10.1109/IMS40175.2024.10600312.

[27] T. E. Djidjekh et al., "Backscattering Rectifier for Security and Identification in the context of Simultaneous Wireless Information and Power Transfer," in 2024 54th European Microwave Conference (EuMC), Paris, France: IEEE, Sep. 2024, pp. 300–303. doi: 10.23919/EuMC61614.2024.10732847.

[28] A. Tepecik and A. F. Ağrak, "Analysis of Lorawan Protocol and Attacks Against Lorawan-Based IoT Devices," ijamec, p. 2, Jun. 2024, doi: 10.58190/ijamec.2024.95.

[29] A. Sidibe, A. Takacs, G. Loubet, and D. Dragomirescu, "Compact Antenna in 3D Configuration for Rectenna Wireless Power Transmission Applications," Sensors, vol. 21, no. 9, p. 3193, May 2021, doi: 10.3390/s21093193.

[30] T. E. Djidjekh, G. Loubet, and A. Takacs, "Backscattering-Based Security in Wireless Power Transfer Applied to Battery-Free BLE Sensors," in 2025 IEEE Wireless Power Technology Conference and Expo (WPTCE), Jun. 2025, pp. 1–4. doi: 10.1109/WPTCE62521.2025.11062233.

[31] X. Gu, R. Khazaka, and K. Wu, "Single-Ended Reconfigurable Wireless Power Harvesting and Harmonic Backscattering," in 2023 IEEE Wireless Power Technology Conference and Expo (WPTCE), San Diego, CA, USA: IEEE, Jun. 2023, pp. 1–4. doi: 10.1109/WPTCE56855.2023.10215912.

[32] V. Palazzi, L. Roselli, M. M. Tentzeris, P. Mezzanotte, and F. Alimenti, "Energy-Efficient Harmonic Transponder Based on On-Off Keying Modulation for Both Identification and Sensing," Sensors, vol. 22, no. 2, p. 620, Jan. 2022, doi: 10.3390/s22020620.

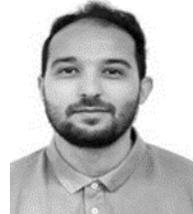

**Taki E. Djidjekh** (Member, IEEE) was born in 1995 in Biskra, Algeria. He received the PhD degree in Electromagnetism and High Frequency Systems in 2025 from INSA Toulouse, France, in collaboration with the LAAS-CNRS Laboratory, Toulouse. He is currently pursuing a postdoctoral position at the LAAS-CNRS Laboratory, Toulouse, France. His research interests include wireless power transfer for IoT applications, as well as the resilience and security of wireless sensor networks. He holds a Master's degree in Electronics of Embedded Systems and Telecommunications from Paul Sabatier University of Toulouse, France (2021), and a Bachelor's degree in Space Telecommunications from the Institute of Aeronautics and Space Studies, Blida University, Algeria (2017).

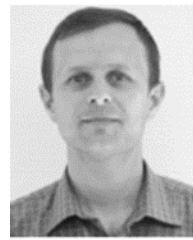

**Alexandru Takacs** (Member, IEEE) was born in Simleu Silvaniei, Romania, in 1975. He received the Engineer Diploma in electronic engineering from the Military Technical Academy, Bucharest, Romania, in 1999, and the master's and Ph.D. degrees in microwave and optical communications from the National

Polytechnic Institute of Toulouse, France, in 2000 and 2004, respectively.

From 2004 to 2007, he was a Lecturer with the Military Technical Academy, and an Associate Researcher with the Microtechnology Institute, Bucharest. From 2008 to 2010, he occupied a postdoctoral position with LAAS-CNRS, Toulouse. During 2011, he was a Research and Development RF Engineer with Continental Automotive SAS France, where he was in charge of antenna design and automotive electromagnetic simulation. Since 2012, he has been an Associate Professor with the University Paul Sabatier, Toulouse, where he performs research within LAAS–CNRS. He has authored or co-authored 5 international patents, 40 papers in refereed journals, one book, one book chapter, and over 120 communications in international symposium proceedings. His research interests include the design of microwave and RF circuits, energy harvesting and wireless power transfer, small antenna design, electromagnetic simulation techniques, and optimization methods.

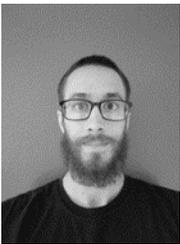

**Gaël Loubet** (Member, IEEE) was born in Toulouse, France, in 1994. He received the Engineer Diploma in Automatic Control and Electronics and the Ph.D. degree in Micro- and Nano-Systems, in 2017 and 2021, respectively, both from INSA Toulouse, France.

From 2021 to 2022, he was a teaching and research associate with the INSA Toulouse and LAAS-CNRS. Since 2022, he has been an Associate Professor with INSA Toulouse and LAAS-CNRS. He has authored more than 30 articles in refereed journals or communications in international conferences. His research interests include electromagnetic Wireless Power Transfer, wireless communications and their low-level security for the Internet-of-Things, Wireless Sensor Networks for Cyber-Physical Systems implementation and Structural Health Monitoring applications, but also the hardware implementation of Artificial Neural Networks in resource-constrained embedded devices.

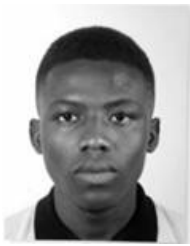

**Lamoussa Sanogo** was born in Mali, in 1995. He is currently a Ph.D. student in the Internet of Things (IoT) physical layer security at the National Institute of Applied Sciences (INSA) of Toulouse, France, since February 2022. His research focuses on finding security solutions and countermeasures adapted to resource-constrained IoT devices. Currently, He is working mainly on a new message authentication technique, based on the polarization shift keying of the radio-frequency signal, for Wireless Sensor Networks (WSN). He is conducting his research activities at the Systems Architecture and Analysis Laboratory (LAAS-CNRS) of Toulouse, France. From 2022 to 2024, he was a teaching associate with the INSA of Toulouse, France.

He received two Master's degrees, the first one in Embedded Systems from the University of Mostaganem, Algeria, in 2018, and the second one in Embedded Systems and Microsystems from the University Paul Sabatier of Toulouse, France, in 2021. He received the bachelor's degree in Electronics from the University of Mostaganem, Algeria, in 2016 His main topics of interest include physical layer security in IoT, reconfigurable antennas, telecommunications, Software-Defined Radio (SDR), signal processing and embedded systems.

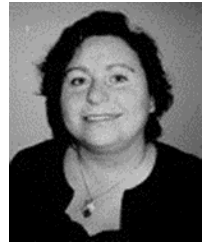

**Daniela. Dragomirescu** (Senior Member, IEEE) is Professor at INSA Toulouse and LAAS-CNRS laboratory. She received the engineering degree with Magna Cum Laude from the Polytechnic University of Bucharest, Romania in 1996, the MSc in circuits design from the University Paul Sabatier, France, in 1997 and the Ph.D. degree with Magna Cum Laude in 2001 from the University of Toulouse, France.

Prof. Dragomirescu is the Deputy Director of LAAS-CNRS laboratory since January 2022. She is the IEEE Solid States Circuits French Chapter Chair. Prof. Daniela Dragomirescu was French Government Fellow of Churchill College, Univeristy of Cambridge in 2014. Daniela Dragomirescu was the Dean of Electrical and Computer Engineering Department at INSA Toulouse since June 2017 to September 2020. Prof. Dragomirescu is conducting research in the area of micro and nano systems for wireless communications with a special focus on Wireless Sensor Networks. She published more than 90 papers in journals and conferences proceedings, 2 patents and she authored 7 academic courses.